\documentstyle[12pt,twoside,fleqn,espcrc1,epsf]{article}
\newcommand{\AmS}{{\protect\the\textfont2
  A\kern-.1667em\lower.5ex\hbox{M}\kern-.125emS}}

\title {\null\vskip -1.0cm\hfill {\small ORNL-CTP-9704 and
hep-ph/9712332}
\\  \vskip 0.8cm 
$J/\psi$ Suppression as a Signal for the Quark-Gluon Plasma
}

\author{Cheuk-Yin Wong\address{Physics Division, Oak Ridge National
Laboratory, Oak Ridge, TN 37831 }%
\thanks{ This research was supported by the
Division of Nuclear Physics, U.S. D.O.E.  under Contract
DE-AC05-96OR22464 managed by Lockheed Martin Energy Research Corp.}
}

\begin{document}
% typeset front matter
\def\bbox{ \vec}
\maketitle

\begin{abstract}

We review the search for the quark-gluon plasma using the signal of
the suppression of $J/\psi$ production in high-energy heavy-ion
collisions.  Recent anomalous $J/\psi$ suppression in high-energy
Pb-Pb collisions observed by the NA50 Collaboration are examined and
compared with earlier results from $pA$ and nucleus-nucleus collisions
with heavy ions of smaller mass numbers.  The anomalous suppression of
$J/\psi$ production in Pb-Pb collisions can be explained as due to the
occurrence of a new phase of strong $J/\psi$ absorption, which sets in
when the number of nucleon-nucleon collisions at a spatial point
exceeds about 4 and corresponds to a local energy density of about 3.4
GeV/fm$^3$.

\end{abstract}

\vskip -0.3cm
\section{Introduction}
\vskip -0.1cm

The study of high-energy heavy-ion collisions is an emerging area of
research (see Ref.\ \cite{Won94} for an introduction).  One of its
objectives is to search for the quark-gluon plasma.  What is a
quark-gluon plasma?  Why do we use high-energy heavy-ion collisions to
search for the quark-gluon plasma.  Why do we use $J/\psi$ production
and suppression as one of the signals for the quark-gluon plasma?
What are the latest experimental results of $J/\psi$ production and
suppression in high-energy heavy-ion collisions?  Why is there
recently new excitement concerning the observation of the anomalous
suppression of $J/\psi$ production in the collision of Pb on Pb at 158
GeV per nucleon?  How do we understand such an anomalous  suppression?

We shall discuss the above-mentioned questions in this brief review.
We shall then turn to examine a model of $J/\psi$ absorption and the
origin of the anomalous suppression in Pb-Pb collisions.

\vskip -0.3cm
\section{  The State of Quark-Gluon Plasma}
\vskip -0.1cm

A quark-gluon plasma is a new state of matter in which quarks and
gluons are deconfined.  This is in distinct contrast to quarks and
gluons in hadron matter where the quarks and gluons of a hadron are
confined within the hadron which has a radial dimension of about a
fermi.  Deconfinement does not mean that quarks and gluons can be
isolated and individually detected.  It only means that quarks and
gluons are allowed to move nearly freely to a greater region of space
outside the radial domain that is usually associated with a hadron.
They are nevertheless still confined within the region of strongly
interacting matter.  For the quark-gluon plasma which may be produced
by high-energy heavy-ion collisions, the transverse dimension of the
plasma is about the size of the overlapping region of nuclear matter
and quarks and gluons can move nearly freely within the distance of a
few fermi, typical of the length of the radii of the colliding nuclei.

A quark-gluon plasma is expected to be the state of lowest energy for
strongly interacting matter at high temperatures or high baryon
densities.  Lattice gauge calculations
\cite{Blu95,Lae96,Kar95,Hat92,Chr92} show that there is a phase
transition between the hadron phase and the quark-gluon plasma phase
at a temperature $T_c$ at which the energy density $\epsilon_c$ is
about $20 T_c^4 $ \cite{Blu95}.  For a pure gauge theory without
fermions, the critical temperature was found to be $265_{-5}^{+10}$
MeV \cite{Lae96}.  With quarks, the critical temperature is
considerably lower.  If one takes 200 MeV to be the order of the
critical temperature $T_c$, the critical energy density $\epsilon_c$ of
the quark-gluon plasma will be of the order of 4 GeV/fm$^3$.

In the lattice gauge theory, the gauge field is described in terms of
the link variable which is the analogue of the spin variable in
ferromagnetism.  We can understand qualitatively that the confined
state of strongly interacting matter at low temperatures arises from
the correlations of the link variables at different spatial locations
at short distances, as well as large distances, so as to minimize the
total energy, resulting in a linear-confining potential between a
quark and an antiquark.  At low temperatures, the tendency to maintain
the correlations between the link variables at different spatial
locations due to the interactions overwhelms the disruptive tendency
due to quantum and thermal fluctuations.  As the temperature
increases, the tendency to disrupt the correlation due to thermal
motion overwhelms the tendency to correlate due to the interactions,
and quarks and gluons become deconfined.

Theoretical work on matter with a high baryon density has not reached
the same degree of sophistication as in the baryon-free case because
of the difficulty of treating dynamical fermions \cite{Hat92,Chr92}.
One expects qualitatively that as the baryon density increases the\break
density of quarks and antiquarks increases, and the pressure due to
degenerate fermions
\vskip -0.0cm \hangafter=-17 \hangindent=-3.2in
\noindent
will increase.  As the baryon density increases above a critical
density, the pressure from the degenerate fermions can overwhelm the
bag pressure of confinement, and there can be a phase transition from
the state of confined quark matter to a state of deconfined
quark-gluon plasma even at zero temperature.  The expected phase
diagram of strongly interacting matter is shown schematically as in
Fig. 1.  Searching and mapping out different parts of the phase
diagram of Fig.\ 1 will be one of the main objectives of high-energy
heavy-ion physics.

\null\vskip 1.0cm \epsfxsize=250pt
\includegraphics{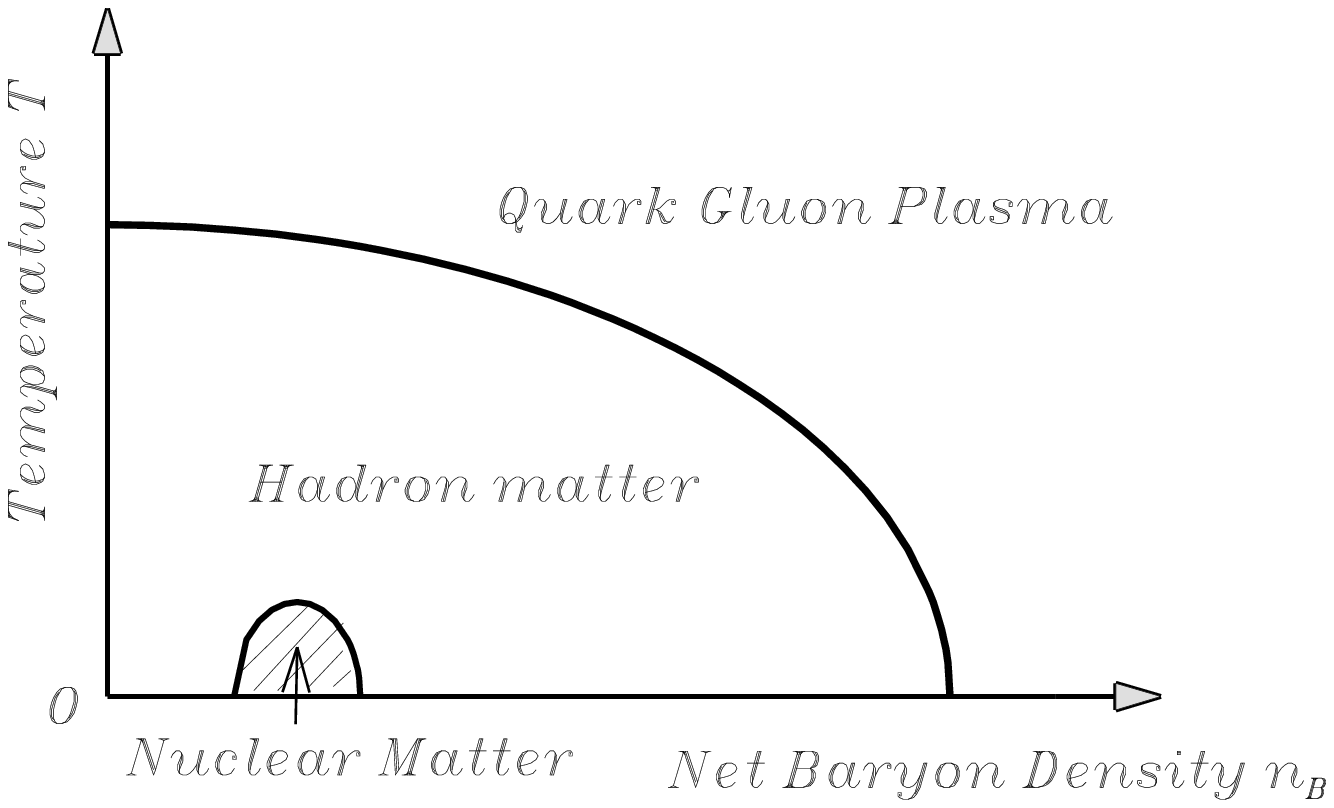}
\vskip -3.3cm
\null\hskip 8.4cm
\begin{minipage}[t]{6.7cm}
\noindent
\bf Fig.1. {\rm { The phase diagram of strongly interacting matter. }}
\end{minipage}
\vskip 4truemm
\noindent 
\vskip -0.4cm

\vskip -0.6cm
\section{  High-Energy Heavy-Ion Collisions}
\vskip -0.1cm

Why do we use high-energy heavy-ion collisions to search for the
quark-gluon plasma?  To appreciate this possibility, it is useful to
take note of the amount of energy that is involved in these
collisions.  We can consider the {Relativistic Heavy-Ion Collider
(RHIC)} which is being constructed at Brookhaven National Laboratory
and will be operational in 1999. The Collider is designed to
accelerate nuclei to an energy of about 100 GeV per nucleon.  In the
collision of a gold nucleus with another gold nucleus in such a
collider, the energy carried by each nucleus is about 100$\times$197
GeV, or 19.7 TeV, and the center-of-mass energy $\sqrt{s}$ is about
2$\times$19.7 TeV, or 39.4 TeV.  The magnitude of energy involved in
nucleus-nucleus collisions is very large indeed.  To attain even
greater center-of-mass energies, there will be a program to accelerate
heavy ions to a center-of-mass energy of about 3 TeV per nucleon in
the {Large Hadron Collider (LHC)} being planned at CERN.  This will
lead to a center-of-mass energy of about 1262 TeV for the collision of
Pb on Pb \cite{Hal95}.

The experimental results in the last two decades indicate that
nucleon-nucleon as well as nucleus-nucleus collisions are highly
inelastic.  Consider first a nucleon-nucleon collision.  In such a
collision, about half of the longitudinal kinetic energy of the
nucleons is lost.  The energy lost is deposited in the vicinity of the
center-of-mass carried away by the quanta of the fields, which are
mostly pions with a small fraction of kaons and baryon pairs.  In a
nucleus-nucleus collision, many nucleon-nucleon collisions take place
and these collisions deposit the energies of the nucleons in the
center-of-mass regions.  In high-energy collisions, because of Lorentz
contraction, all the nucleon-nucleon collisions take place at about
the same time in close spatial proximity in the nucleon-nucleon
center-of-mass system.  As a result, a region of very high energy
density, as high as a few GeV per cubic fermi, will be produced.  Such
an energy density is within the same order as that expected to be the
critical energy density leading to the transition from the hadron
matter to the quark-gluon plasma.  The temperature of the matter in
the central region near the center-of-mass is also high.  It is
therefore reasonable to look for the quark-gluon plasma using
high-energy heavy-ion collisions.
\vskip -1cm
\null{}
\vskip -2.8cm
\section{  Signals for the Quark-Gluon Plasma}
\vskip -0.1cm

In a high-energy heavy-ion collision, one starts with a cold nuclear
matter in collision.  As a result of the collision, a quark-gluon
plasma may be formed.  The quark-gluon plasma can only be a 
transient state of matter, as the matter will cool down and will
return to the hadron phase at the end. It is during the excursion into
the quark-gluon plasma phase that  it will leave the signals of the
quark-gluon plasma.

There are generally two different methods.  The first one is to study
the reaction products of the constituents of the quark-gluon plasma.
For example, one can use $gq \rightarrow \gamma q$, $g\bar q
\rightarrow \gamma \bar q$, and $q\bar q \rightarrow \gamma g$ and
look at the momentum distribution of the product gamma particles to
infer the momentum distribution of the gluons and quarks in the plasma
\cite{Kap92}.  One can also use the dilepton spectrum in the reactions
of $gq \rightarrow l^+ l^- q$, $g\bar q \rightarrow l^+l^- \bar q$,
and $q\bar q \rightarrow l^+l^- g$ in order to infer the momentum
distributions of the gluons and quarks in the plasma \cite{Kaj86}.

In the second method, one makes use of the peculiar properties of the
plasma to infer the existence of the plasma.  One can study the
suppression of $J/\psi$ production \cite{Mat86}, the enhancement of
strangeness production \cite{Ref82}, the dependence of the energy
density as a function of the temperature \cite{Van85}, the restoration
of chiral symmetry\cite{Blu95,Lae96,Kar95,Hat92,Chr92}, or the large
effective lifetime of the system \cite{Ham88}.

In using any of the above signals to study the quark-gluon plasma,
there needs to be a careful subtraction of signals from other sources.
Such a subtraction is not without ambiguities and uncertainties.
Thus, a clear identification of the quark-gluon plasma will require
not just a single piece of data but rather many collaborative pieces
of evidence pointing to peculiarities associated with the quark-gluon
plasma.

\vskip -0.3cm
\section{ Suppression of  $J/\psi$  Production as a Signal for the Quark-Gluon
Plasma}

$J/\psi$ is a bound state consisting of a $c$ quark and a $\bar c$
quark.  In free space, the $c$ and $\bar c$ interact through a linear
confining potential and a color-Coulomb interaction.  In high-energy
heavy ion collisions, $J/\psi$ may be produced in one of the many
nucleon-nucleon collisions.  If a quark-gluon plasma is formed in the
meantime and this $J/\psi$ finds itself in the environment of the
quark-gluon plasma, then two things will happen.  First, the linear
confining potential between the charm quark and the charm antiquark
will not be present because the temperature of the quark-gluon plasma
brings the $J/\psi$ particle to the state of deconfinement. Second,
the color-Coulomb interaction between the charm quark and the charm
antiquark will be modified because the light quarks, antiquarks, and
gluons of the plasma can move around from other locations to screen
the charm quark color charge from the charm antiquark.  This reduces
the color charge seen by the charm quark with respect to the charm
antiquark.  The color-Coulomb interaction is modified into a
color-Yukawa interaction, with the range of the Yukawa depending
inversely on the temperature.  Such an inverse dependence arises
because the greater the temperature , the greater the quark density
surrounding the charm quark, and the greater is the degree of
screening to diminish the interaction between the charm quark and the
charm antiquark.

As the range of the color-Yukawa interaction decreases, the energy
levels of the $c\bar c$ system rise.  In other words, the energy level
of the $c\bar c$ system rises with temperature.  When the temperature
rises to a critical temperature, there can be a resonance of $c$ and
$\bar c$ at $E=0$ \cite{Won97zp}. As the temperature rises even
higher, then there will be no bound state between $c$ and $\bar c$,
and no matter how close the $c$ is brought to $\bar c$, the $c$ and
$\bar c$ will drift apart.  When the phase of the quark-gluon plasma
has passed, the $c$ and $\bar c$ will find themselves so far apart
that the formation of a bound state is no longer possible, and the
charm quark will pick up a light antiquark, while the charm antiquark
will pick up a light quark, to form $D$ and $\bar D$ states.  They
will not be part of the bound state formation process.  They are lost
from bound state production, and thus in the presence of the
quark-gluon plasma, the production of $J/\psi$ is suppressed
\cite{Mat86}.

\vskip -0.3cm
\section{ Advantages and Disadvantages of Using $J/\psi$
Suppression as a Signal for the Quark-Gluon Plasma}
\vskip -0.1cm

One can select the decay product of $J/\psi$ to be dileptons and look
for dileptons which carry the signature of $J/\psi$.  As leptons
interact only electromagnetically with hadron matter and the
interaction is not strong, the leptons are not affected by the
final-state interaction due to the presence of the strongly
interacting matter.  Thus, one advantage of $J/\psi$ as a signal of
the quark-gluon plasma is that the detection of $J/\psi$ through
dilepton coincidence is less subject to final-state interactions. The
dileptons carry information on the system at the moment of their
creation.  Furthermore, the calibration of the dilepton cross sections
can be made reliably by comparing with dilepton production through the
Drell-Yan process, which arises from the collision of the quark parton
of one nucleon with the antiquark parton from the other nucleon and is
quantitatively very well understood.

The disadvantage of $J/\psi$ is that $J/\psi$ is also suppressed by
non-quark-gluon plasma sources, and these must be taken into account
to get information on the suppression by the quark-gluon plasma.  In
addition, the production and absorption mechanism for $J/\psi$ with
low transverse momenta at a few hundred GeV is not completely
understood.  Much theoretical and experimental work is needed to
construct a picture of $J/\psi$ production and absorption.
\vskip -0.3cm

\vskip -1.6cm
\section{  Experimental $J/\psi$ Suppression Data} 
\vskip -0.1cm

Experimental data indeed show that $J/\psi$ production in
nucleus-nucleus collisions is suppressed relative to $pp$ collisions.
In Figs. 2$a$ and 2$b$, we plot ${\cal B}\sigma(A+B\rightarrow J/\psi
X)/AB$ and ${\cal B}'\sigma(A+B\rightarrow \psi' X)/AB$ with $x_F>0$
as a function of $\eta= A^{1/3}(A-1)/A + B^{1/3}(B-1)/B $, where
$\eta$ is proportional to average path length $L$ through which a
produced $J/\psi$ needs to pass in order to come out of the nuclear
matter, and $\cal B$ and ${\cal B}'$ are respectively the branching
ratio for $J/\psi$ and $\psi'$ to decay into $\mu^+\mu^-$.  If there
were no absorption, ${\cal B} \sigma(A+B\rightarrow J/\psi X)/AB$
would be independent of the path length $L$ or the parameter $\eta$.
However, for $pA$ collisions, the yield of $J/\psi$ per
nucleon-nucleon collision depends on the path length in a simple
exponential way as shown in Fig.\ 2$a$.  We plot the similar yield of
$\psi'$ as a function of $\eta$ in Fig.\ 2$b$.

\vskip -0.5cm
\epsfxsize=400pt
\includegraphics{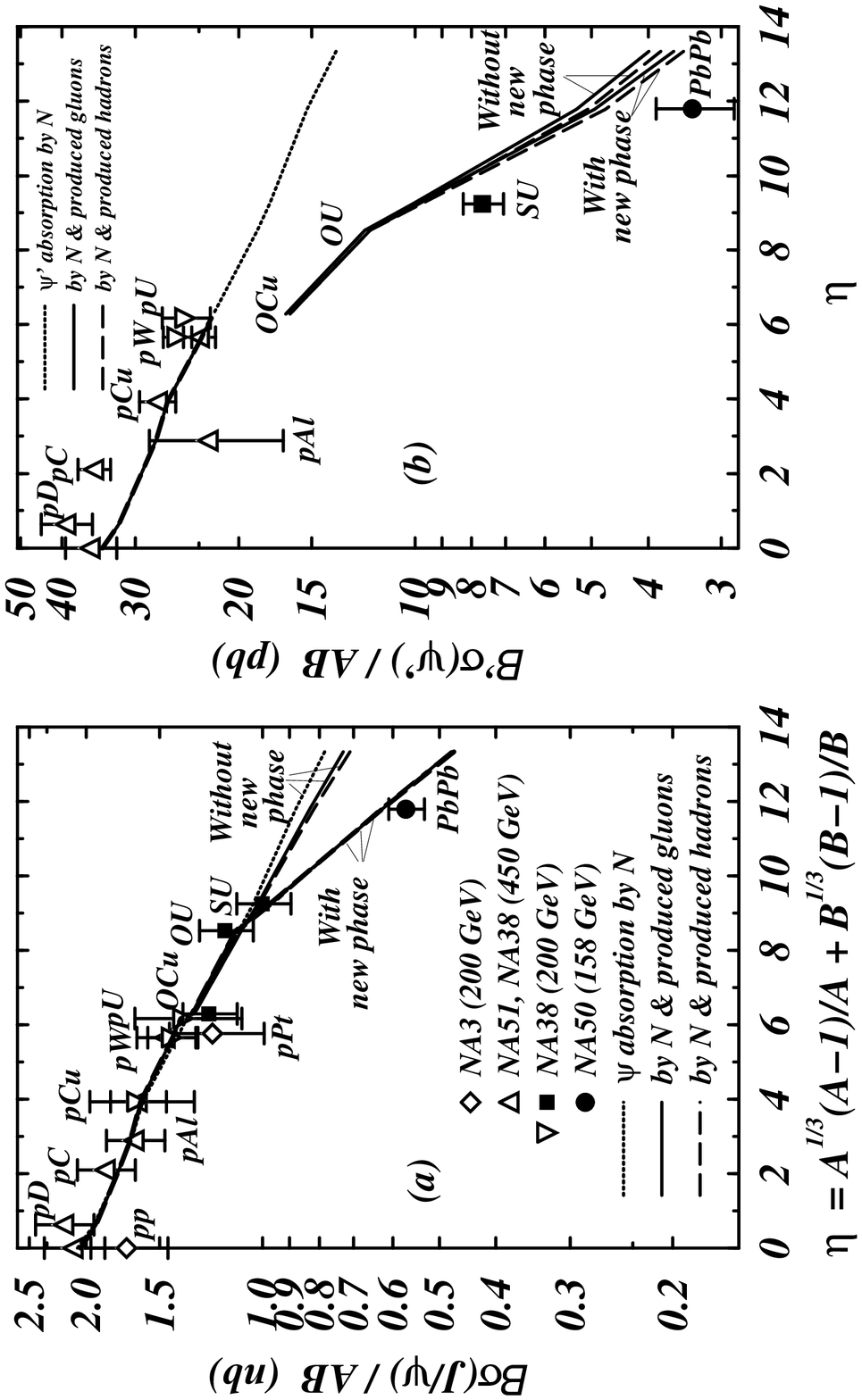}
\vskip 10cm
\begin{minipage}[t]{6in}
\noindent \bf Fig.2.  \rm
{ ($a$) \protect${\cal B}\sigma_{J/\psi}^{AB}/AB\protect$ and
($b$) \protect${\cal B}'\sigma_{\psi'}^{AB}/AB\protect$ as a function
of $\eta$.  Data are from NA3 \protect\cite{Bad83}, NA51
\protect\cite{Bal94}, NA38 \protect\cite{Bag89,Lou95,Bag95}, and NA50
\protect\cite{Gon96,Lou96}.  }
\end{minipage}
\vskip 4truemm
\noindent 

As was recognized early by Andersen et al. \cite{And77} and by Gershel
and H\" ufner \cite{Ger88}, Fig.\ 2$a$ shows that $J/\psi$ is also
suppressed by non-quark-gluon-plasma sources, which must be taken into
account to get information on the suppression by the quark-gluon
plasma.

\vskip -0.3cm
\section{  $J/\psi$ Absorption by Collision with Nucleons}
\vskip -0.1cm

In a nucleon-nucleon collision, $c\bar c$ pairs are produced by the
collision of a parton of one nucleon with a parton of another nucleon.
Among the $c\bar c$ pairs, only those with low invariant masses will
evolve into precursors of $J/\psi$ and $\psi'$.  Because $J/\psi$ are
produced predominantly at rapidity zero, the precursor is produced as
essentially at rest in the center-of-mass system, and the fate of the
precursor can be studied best in the nucleon-nucleon center-of-mass
frame. The precursor will encounter projectile and target nucleons at
high relative energies. The absorption by collisions with these
nucleons constitutes the hard component of absorption which is present
in both $pA$ and (nucleus $A$)-(nucleus $B$) collisions.

It has been suggested recently that the produced precursor exists as a
coherent admixture of color and angular momentum states \cite{Won97c}.
The total cross section for this coherent precursor for a collision
with a nucleon at high energy can be evaluated by the two-gluon
model of the Pomeron \cite{Low75,Nus75,Dol92}.  The total cross
section is approximately the weighted sum of the color-singlet and
color-octet total cross sections. As the color-singlet cross section
is very small, and the color-octet cross section is very large (of the
order of 30 to 60 mb), the absorption cross section depends on the
color-octet fraction.  The experimental absorption cross section
suggests a color-octet fraction of the order of 20\%.  Different
parton combinations will lead to different color admixtures and
different absorption cross sections.

In a situation where the precursor comes predominantly from a single
combination of partons, as in the present case when $gg$ fusion is the
dominant process for $c\bar c$ quarkonium production at fixed target
energies, the precursor state $\Psi_{gg}(L)$ is absorbed in its
passage through nuclear matter by a single precursor-nucleon cross
section $\sigma_{abs}$. The survival probability for the precursor
after traveling a path length $L$ in nuclear matter is
\begin{eqnarray}
|\Psi_{gg}(L) > = e^{-\rho \sigma_{abs} L/2} |\Psi_{gg}(L=0)>\,,
\end{eqnarray}
with the survival probability $e^{-\rho \sigma_{abs} L}$, where $\rho$
is the number density of nucleons in a nucleus at rest.  Because the
production of different bound states comes from the projection of the
coherent precursor state onto the bound states after the absorption,
the production of various $c\bar c$ bound states in the nuclear medium
are suppressed by the same rate with the same $\sigma_{abs}$.  Thus,
in $pA$ collisions where the precursors are absorbed only by
collisions with nucleons, we expect that the survival factors for the
production of various $c\bar c$ bound states are characterized by
approximately the same absorption cross section $\sigma_{abs}$.

The approximate equality of the absorption cross sections for $J/\psi$
and $\psi'$ production in $pA$ collisions is indeed observed
\cite{Lou95,Bag95,Ald91}, as one can see from the approximate equality
of the slope of the $pA$ lines for $J/\psi$ and $\psi'$ in Figs.\ 2$a$
and 2$b$.  This is further collaborated by another piece of data in
Figs.\ 3$a$ and 3$b$ where we plot respectively the ratio ${\cal
B}(J/\psi) \sigma(J/\psi)/\sigma({\rm Drell-Yan})$ and ${\cal
B}'(\psi') \sigma(\psi')/\sigma({\rm Drell-Yan})$.  Because the
Drell-Yan cross section does not suffer much absorption in hadron
environments, the ratios of ${\cal B}(J/\psi) \sigma(J/\psi)$ and
${\cal B}(\psi') \sigma(\psi')$ to the Drell-Yan cross section are
proportional to $J/\psi$ and $\psi'$ survival probabilities.  These
ratios, as a function of the path length $L$, are presented for
$J/\psi$ in Fig. 3$a$ and for $\psi'$ in Fig. 3$b$.  The slopes of the
$pA$ lines for $J/\psi$ and $\psi'$ are nearly the same.  Therefore
for the hard component of the absorption process, the rate of
absorption and the associated absorption cross sections by collisions
with nucleons at high energies are approximately the same for $J/\psi$
and $\psi'$.

\null\vskip -1.0cm
\epsfxsize=400pt
\includegraphics{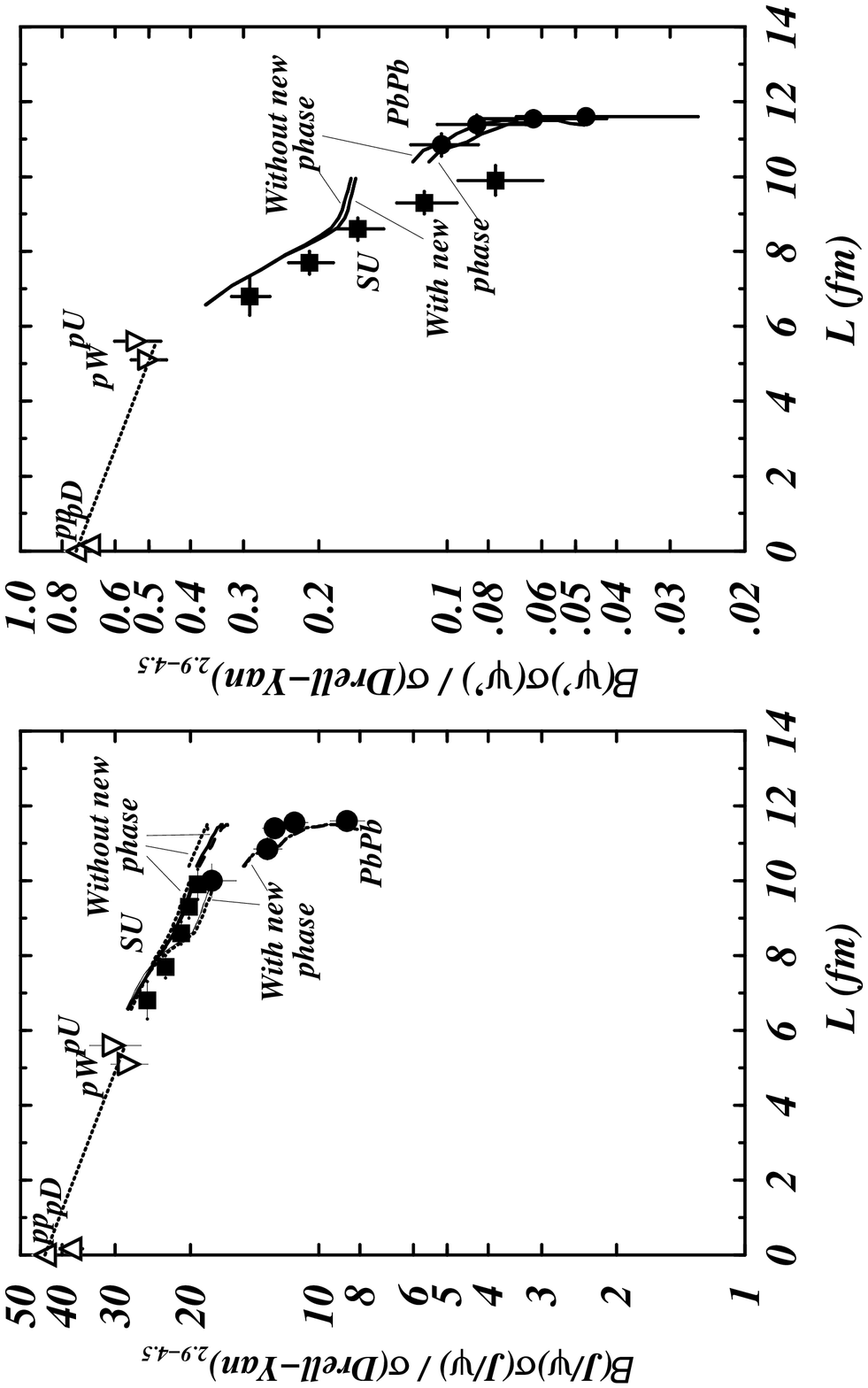}
\null\vskip 9.2cm
\begin{minipage}[t]{6in}
\noindent \bf Fig.3.  \rm
{ ($a$) \protect${\cal
B}\sigma_{J/\psi}^{AB}/AB\protect$ and ($b$) \protect${\cal
B}'\sigma_{\psi'}^{AB}/AB\protect$ as a function of the path length
$L$. Data are from NA51
\protect\cite{Bal94}, NA38 \protect\cite{Bag89,Lou95,Bag95}, and NA50
\protect\cite{Gon96,Lou96}. }
\end{minipage}
\vskip 4truemm
\noindent

\vskip -2.0cm
\section{  Absorption of $J/\psi$ and $\psi'$ Precursors by Produced
Soft Particles}
\vskip -0.1cm

An additional absorption component is present in nucleus-nucleus
collisions where a $J/\psi$ or $\psi'$ precursor can find itself in
the middle of fireballs of soft particles produced by earlier or later
nucleon-nucleon collisions centered at the same spatial location.
These produced soft particles may exist in the form of virtual gluons
in early stages and hadrons at later stages \cite{Web84}.  They will
collide with the $J/\psi$ or $\psi'$ precursor to lead to its breakup.
The centers of the fireballs of produced soft particles are nearly at
rest, and the collisions between the precursor and the soft particles
occur at a kinetic energy which is about the fireball temperature (of
a few hundred MeV).  Absorption of the precursor due to the collisions
with soft particles constitutes the soft component of absorption.

The survival probability of $J/\psi$ and $\psi'$ due to the absorption
by soft particle collisions is approximately an exponential function
whose exponent is proportional to the (precursor)-(soft particle)
absorption cross section and the density of soft particles.  The
number of produced soft particles depends on the number of
participants \cite{Sor88}.  The number of participants in turn is
proportional to the longitudinal path length passing through nuclei
$A$ and $B$.  Because of such dependence on participant numbers and
the longitudinal path length, the survival probability due to soft
particle absorption is approximately $\exp\{ -c \sigma({\rm
precursor}$-(soft particle))$ L\}$, where $c$ is approximately a
constant (see pages 374-377 of Ref. \cite{Won94}).

How does $\sigma({\rm precursor}$-(soft particle)) depend on the
nature of the produced particles?  The collision of the precursor with
the soft particle at low energies brings up a different type of
interaction between them. Because these collisions take place at low
energies, the natural basis states to describe the interaction are the
$c\bar c$ quarkonium states, and the interaction depends on the
separation energy of the bound $c\bar c$ quarkonium states. Thus, at
such a low energy precursor-(soft particle) collision, one projects
out the precursor state into various bound states, and the absorption
cross section for each component depends on the separation energy of
the bound states.  The energy required to break up a $J/\psi, \chi_1,
\chi_2$ and $\psi'$ is 640 MeV, 228 MeV, 182 MeV, and 52 MeV
respectively.  On the other hand, the relative energy between the
precursor and soft particles is approximately the temperature of the
soft particles, which is about 150 to 200 MeV.  Thus, the cross
section for the breakup of $\psi'$ is much greater than the cross
section for the breakup of $J/\psi$ and $\chi$'s.  The rate of
$J/\psi$ and $\chi$ absorption by soft particle collisions is much
less than the rate of absorption of $J/\psi$.

To study the soft component of absorption, one looks for a gap and a
change of the slope in going from the $pA$ line to the $AB$ line.
From the $\psi'$ data in Fig. 2$b$, one can discern the presence of a
gap in going from the $pA$ line to the S-U data point.  For the
nucleus-nucleus collision data in Fig.\ 3, one makes use of the
information on the transverse energies to select nucleus-nucleus
collisions with different impact parameters and different average path
lengths $L$.  The $\psi'$ data in Fig. 3$b$ show a big gap and a large
change of the slope in going from the $pA$ line to the $AB$ line,
conforming to the signature of the soft component.  Figs.\ 2$b$ and
3$b$ indicate that for $\psi'$ production, the absorption due to the
soft component is large.

The above analysis of the $\psi'$ data provides us with a clear
signature of the soft absorption component.  This signature can be
used to identify the soft component in other production processes.
Upon searching for the signature of the soft absorption component in
$J/\psi$ production for $pA$, O-Cu, O-S, and S-U collisions in
Figs. 2$a$ and 3$a$, one finds that there is almost no gap and no
change of the slope in going from the $pA$ line to the $AB$ line.  One
concludes that the absorption of $J/\psi$ by soft particles, as
revealed by the data of $pA$, O-Cu, O-U, and S-U, is small.

When one extends one's consideration to Pb-Pb collisions, one finds
that the $AB$ line of O-Cu, O-U, and S-U in Figs.\ 2$a$ and 3$a$ are
much above the Pb-Pb data points.  This indicates that the degree of
$J/\psi$ absorption by soft particles, as revealed by the data of
O-Cu, O-U, and S-U collisions, cannot explain the Pb-Pb data, and a
new phase of strong $J/\psi$ absorption in Pb-Pb collisions is
suggested \cite{Won96a,Won96qm,Won97prc}.  Similar conclusions have
been reached also by other workers \cite{Bla96qm,Kha96qm}.
\null\vskip -2.1cm
\section{Microscopic Absorption Model of $J/\psi$ and $\psi'$ Absorption}
\vskip -0.1cm

We can be more quantitative to study this departure of Pb-Pb data by
using the microscopic absorption model of
\cite{Won96a,Won96qm,Won97prc}.  In this model, each nucleon-nucleon
collision is a possible source of $J/\psi$ and $\psi'$ precursors.  It
is also the source of a fireball of soft particles which can absorb
$J/\psi$ and $\psi'$ precursors produced by other nucleon-nucleon
collisions.  One follows the space-time trajectories of precursors,
baryons, and the centers of the fireballs of soft particles.
Absorption occurs when the space-time trajectories of the precursors
cross those of other particles.  Using a row-on-row picture in the
center-of-mass system and assuming straight-line space-time
trajectories, we obtain the differential cross section for $J/\psi$
production in an $AB$ collision as \cite{Won96a}
\begin{eqnarray}
\label{eq:fin}
{ d \sigma_{{}_{J/\psi}}^{{}^{AB}} ({\bf b}) \over \sigma_{{}_{J/\psi
}}^{{}^{NN}}~d{\bf b} } &=& \!\! \int \!\! { d{ {\bf b}}_{{}_A} \over
\sigma_{\rm abs}^2(J/\psi- N) } \biggl \{ 1 -\biggl [ 1-
T_{{}_A}({\bf b}_{{}_A}) \sigma_{\rm abs}(J/\psi-N) \biggr ] ^A \biggr
\} \nonumber\\
& &~~~~~~~\times \biggl \{ 1 -\biggl [ 1- T_{{}_B}({\bf b}-{\bf b}_{{}_A})
\sigma_{\rm abs}(J/\psi- N) \biggr ] ^B \biggr \} F({\bf b}_A, {\bf b})\,,
\end{eqnarray}
where $T_A(\bf{b}_A)$ is the thickness function of nucleus $A$, and
$F({\bf b}_A, {\bf b})$ is the survival probability due to soft
particle collisions.  To calculate $F({\bf b}_A, {\bf b})$, we sample
the target transverse coordinate ${\bf b}_A$ for a fixed impact
parameter ${\bf b}$ in a row with the nucleon-nucleon inelastic cross
section $\sigma_{in}^{NN}$.  In this row, $BT_B({\bf b}-{\bf
b}_A)\sigma_{in}$ projectile nucleons will collide with $AT_A({\bf
b}_A)\sigma_{in}$ target nucleons.  We construct the space-time
trajectories of these nucleons to locate the position of their
nucleon-nucleon collisions.  These collisions are the sources of
$J/\psi$ and $\psi'$ precursors and the origins of the fireballs of
produced particles.  For each precursor source from the collision $j$
and each absorbing fireball from the collision $i$ at the same spatial
location, we determine the time when the precursor source coexists
with the absorbers as virtual gluons $t_{ij}^g$ or as produced hadrons
$t_{ij}^h$.  The survival probability due to this combination of
precursor source and absorber is then $\exp \{-k_{\psi g} t_{ij}^g
-k_{\psi h} t_{ij}^h \}$, where the rate constant $k_{\psi m}$ for
$m=g, h$ is the product of $\sigma_{\rm abs}(J/\psi$-$ m)$, the
average relative velocity $v_m$, and the average number density
$\rho_m$ per $NN$ collision.  When we include all possible precursor
sources and absorbers, $F({\bf b}_A,{\bf b})$ becomes
\begin{eqnarray}
\label{eq:fb}
F({\bf b}_A, {\bf b}) \!= \!\sum_{n=1}^{N_<}  \!{ a(n)\over {N}_> {N}_<} 
\!\sum_{j=1}^n
\!\exp\{- \theta \!\!\!\! \sum_{i=1, i\ne j}^n 
\!\!\!( k_{\psi g} t_{ij}^g +
k_{ \psi h} t_{ij}^h ) \} \,,
\end{eqnarray}
where $N_>({\bf b}_A)$ and $N_<({\bf b}_A)$ are the greater and the
lesser of the (rounded-off) nucleon numbers $AT_A({\bf
b}_A)\sigma_{in}$ and $BT_B({\bf b}-{\bf b}_A)\sigma_{in}$, $a(n)=2
{\rm~~for~~} n=1,2,...,N_<-1 $, and $ a(N_<)=N_>-N_<+1$.  The function
$ \theta$ is zero if $A=1$ or $B=1$ and is 1 otherwise.  The survival
probability $F$ can be determined from plausible $c\bar c$, $g$, $h$
production time $t_{c\bar c}$, $t_g$, $t_h$, and the freezeout time
$t_f$ \cite{Won96a}. The cross section for $\psi'$ production can be
obtained from the above equations by changing $J/\psi$ into $\psi'$.

We use this microscopic absorption model to study the experimental
data. Consider first the $J/\psi$ data in $pA$, O-Cu, O-U, and S-U
collisions.  If one assumes that there is no soft particle absorption,
the results with the least $\chi^2$ are obtained with $\sigma_{\rm
abs}(J/\psi$-$N)=6.94$ mb, shown as dotted curves in Figs. 2$a$ and
3$b$.  If one assumes additional absorption by soft particles in the
form of produced gluons or produced hadrons, one obtains fits shown
respectively as the solid and dashed curves marked by ``without new
phase'' in Figs. 2$a$ and 3$a$.  The results indicate that the soft
component of $J/\psi$ absorption, as revealed by O-Cu, O-U, and S-U
collisions, is small, and the extrapolated results from any one of
these three descriptions are much greater than the Pb-Pb data points.
The Pb-Pb data cannot be explained by the absorption due to collisions
with nucleons and soft particles.

We next examine the $\psi'$ data in Figs. 2$b$ and 3$b$.  The
theoretical results with no soft particle absorption are given by the
dotted curves, which fit the $pA$ data, but are much too large for the
S-U data.  Theoretical results calculated with
$\sigma(\psi'$-$N)=6.36$ mb with additional absorption by produced
gluons or produced hadrons are shown respectively as the solid and
dashed curves marked by ``without new phase'' in Figs. 2$a$ and 3$a$.
These theoretical results indicate that soft particle absorption leads
to a gap and a change of the slope when one goes from the $pA$ line to
the $AB$ line.  A large soft component is needed to explain the
$\psi'$ data in S-U collisions.  The flattening of the theoretical
ratio of $B(\psi')\sigma(\psi')/\sigma({\rm Drell-Yan})$ as a function
of $L$ for S-U collisions in Fig. 3$b$ arises because the distribution
of soft particle densities in the central region of $\psi'$ production
is insensitive to the impact parameter for small impact parameters
when the masses of the two colliding nuclei are very different.
\vskip -0.1cm

\vskip -2.0cm
\section{New Phase of $J/\psi$ Absorption}
\vskip -0.1cm

The deviation of the $J/\psi$ data in Pb-Pb collisions from the
conventional theoretical extrapolations in $p$-A, O-A, and S-U
collisions suggests that there is a transition to a new phase of
strong absorption which sets in when the local energy density exceeds
a certain threshold.  We can extend the absorption model to describe
this transition.  The energy density at a particular spatial point at
a given time is approximately proportional to the number of
nucleon-nucleon collisions which has taken place at that spatial point
up to that time.  We use the row-on-row picture as before, and
postulate that soft particles make a transition to a new phase with
stronger $J/\psi$ absorption characteristics at a spatial point if
there have been $N_c$ or more baryon-baryon collisions at that spatial
point.  The quantity $k_{\psi g} t_{ij}^g + k_{\psi h} t_{ij}^h$ in
Eq.\ (\ref{eq:fb}) becomes $k_{\psi g} t_{ij}^g + k_{\psi h} t_{ij}^h
+ k_{\psi x} t_{ij}^x$.  Here, the new rate constant $k_{\psi x}$
describes the rate of absorption of $J/\psi$ by the produced soft
matter absorber in the new phase. Also, the quantity $t_{ij}^x$ is the
time for a $J/\psi$ produced in collision $j$ to coexist at the same
spatial location with the absorbing soft particles produced in
collision $i$ in the form of the new phase, before hadronization takes
place.  Baryons passing through the spatial region of the new phase
may also become deconfined to alter their $J/\psi$ absorption
characteristics.  Accordingly, we also vary the effective absorption
cross section, $\sigma_{\rm abs}^x(\psi N)$, for $\psi$-$N$
interactions in the row in which there is a transition to the new
phase, while the absorption cross section $\sigma_{\rm abs}(\psi N)$
remains unchanged in other rows where there is no transition.  As
shown on the curves marked by ``with new phase'' in Figs.\ 2$a$ and
3$a$, model results assuming a new phase give good agreement with
${\cal B}\sigma_{J/\psi}^{AB}/AB$ data including the Pb-Pb data point,
with the parameters $N_c=4$, $k_{\psi x}=1$ c/fm.  The rate constant
$k_{\psi x}$ for this new phase is much greater than the corresponding
rate constants $k_{\psi g}$ or $k_{\psi h}$.

We can study the $\psi'$ data to see how the presence of the new phase
will affect $\psi'$ production.  Theoretical results obtained by
assuming the new phase are shown as the curves marked by ``with new
phase'' in Fig. 2$b$ and the lower solid curves in Fig. 3$b$.  They
indicate that as far as $\psi'$ suppression is concerned, the $\psi'$
particles are so strongly absorbed by collisions with soft particles
that the presence of an additional source of absorption leads only to
a very small additional absorption.  Seen in this light, $J/\psi$ is a
better probe for a new phase of absorption than $\psi'$ because of its
large threshold value which does not allow it to be destroyed in great
proportion by soft particles.

We have seen that the anomalous suppression of $J/\psi$ in Pb-Pb
collisions can be explained by models in which a new phase of strong
absorption sets in when the number of baryon-baryon collisions at a
local point exceeds or is equal to $N_c=$4.  We can inquire about the
approximate threshold energy density $\epsilon_c$ which corresponds to
the onset of the new phase.  Evaluated in the nucleon-nucleon
center-of-mass system, the energy density at the spatial point which
has had $N_c$ prior nucleon-nucleon collisions can be estimated by
knowing the average number of produced particles per nucleon-nucleon
collision, the average energy carried by each particle, and the
spatial separation between adjacent collisions.  For $N_c=4$ we find
$\epsilon_c \sim 3.4$ GeV/fm$^3$, which is close to the quark-gluon
plasma energy density, $\epsilon_c \sim 4.2$ GeV/fm$^3$, calculated
from the lattice gauge theory result of $\epsilon_c/T_c^4 \sim 20$
\cite{Blu95} with $T_c \sim 0.2$ GeV.  Therefore, it is interesting to
speculate whether the new phase of strong absorption may be the
quark-gluon plasma.  In an equilibrated or non-equilibrated
quark-gluon plasma, the screening of the $c$ and $\bar c$ quarks by
deconfined quarks and deconfined gluons will weaken the interaction
between $c$ and $\bar c$ and will enhance the breakup probability of a
quasi-bound $(c \bar c)$ system.
\vskip -0.0cm

\vskip -0.6cm
\section{ Conclusion and Discussions}
\vskip -0.1cm

$J/\psi$ and $\psi'$ precursors produced in high-energy heavy-ion
collisions are absorbed by their collisions with nucleons and produced
soft particles, leading to two distinct absorption mechanisms.
The absorption by nucleons occurs at high kinetic
energies between the precursor and the nucleon and constitutes the
hard component of absorption.  It is operative in $pA$ and $AB$
collisions.  The absorption by soft particles occurs at low relative
energies, and constitutes the soft component of absorption.  It occurs
mainly in $AB$ collisions.

The signature for the hard component is an approximately straight line
in the semi-log plot of the survival probability for $pA$ collisions
as a function of the path length or $\eta= A^{1/3}(A-1)/A +
B^{1/3}(B-1)/B $.  The slope of the line gives the precursor-nucleon 
absorption cross section.

The signature for the soft component consists of a gap and a change of
the slope from the $pA$ line to the $AB$ line in the semi-log plot of
the survival probability as a function of the path length or $\eta$.
The greater the gap, the greater the change of the slope, and vice
versa.  Application of these signatures to examine the $J/\psi$ and
$\psi'$ data indicates that the degree of absorption by soft particles
on $J/\psi$ production, as revealed by the O-Cu, O-U, and S-U data, is
small.  The absorption by soft particles on $\psi'$ production is,
however, quite large.

A microscopic absorption model which takes care of all precursor
sources and absorbers is used to examine these two mechanisms.  The
microscopic model results support the above qualitative descriptions.

When one extends one's consideration to Pb-Pb collisions, one finds
that the degree of $J/\psi$ absorption by soft particles as
constrained by the data of O-Cu, O-U, and S-U collisions cannot
explain the Pb-Pb data, and a new phase of strong $J/\psi$ absorption
in Pb-Pb collisions is suggested.  The anomalous suppression of
$J/\psi$ production in Pb-Pb collisions can be explained as due to the
occurrence of a new phase of strong $J/\psi$ absorption, which sets in
when the number of nucleon-nucleon collisions at a spatial point
exceeds about 4 and corresponds to a local energy density of about 3.4
GeV/fm$^3$.  

In order to demonstrate that the anomalous suppression in Pb-Pb
collisions arises from the occurrence of the quark-gluon plasma, it is
necessary to obtain additional pieces of collaborative experimental
evidence.  Much future work remains to be done to identify the
quark-gluon plasma if it is ever produced in high-energy Pb-Pb
collisions.

%\newpage

\end{document}